\documentclass[11 pt]{article}
\usepackage[legalpaper, margin=1.0in]{geometry}
\usepackage[round]{natbib}
\usepackage[utf8]{inputenc}
\usepackage{amsmath, amssymb,amsfonts,amsthm}
\usepackage{graphicx} 
\usepackage{apptools}
\usepackage{url}
\def\a{\mathbf{a}}

\def\e{\mathbf{e}}
\def\y{\mathbf{y}}
\def\z{\mathbf{z}}
\def\A{\mathbf{A}}

\def\x{\mathbf{x}}
\def\y{\mathbf{y}}

\def\L{\mathcal{L}}

\def\Y{\mathbf{Y}}
\def\real{\mathcal{R}}
\DeclareMathOperator*{\argmin}{argmin}
\newcommand{\vectornorm}[1]{\left|\left|#1\right|\right|}
\newtheorem{theorem}{Theorem}

\newtheorem{definition}{Definition}
\author{Abiy Tasissa\thanks{Department of Mathematics, Tufts University, Medford, MA 02155, USA.} \and Pranay Tankala \thanks{School of Engineering and Applied Sciences, Harvard University, Cambridge, MA 02138, USA.} 
\and Demba Ba\footnotemark[2]
}

\begin{document}
\title{Weighed $\ell_1$ on the simplex: Compressive sensing meets locality}
\date{\vspace{-5ex}}


\maketitle
\begin{abstract}
Sparse manifold learning algorithms combine techniques in manifold learning and sparse optimization to learn features that could be utilized for downstream tasks. 
The standard setting of compressive sensing can not be immediately applied to this setup. Due to the intrinsic geometric structure of data, dictionary atoms
might be redundant and do not satisfy the restricted isometry property or coherence condition. In addition, manifold learning emphasizes learning local geometry
which is not reflected in a standard $\ell_1$ minimization problem. We propose weighted $\ell_0$ and weighted $\ell_1$ metrics that encourage representation
via neighborhood atoms suited for dictionary based manifold learning. Assuming that the data is generated from Delaunay triangulation, we show the equivalence of weighted $\ell_0$ and weighted $\ell_1$. We discuss an optimization program that learns the dictionaries and sparse coefficients and demonstrate the utility of our regularization on synthetic and real datasets. 

               \end{abstract}

\section{Introduction}
The compressive sensing (CS) problem considers the recovery of a sparse vector $\x \in \real^{m}$ given $d$ undetermined measurements $\y = \A\x$. The optimization problem is given by 
\begin{equation}\label{eq:l0_min}
\underset{\x \in \real^{m}}{\min}\quad \vectornorm{x}_{0} \quad \text{s.t.} \quad \y = \A\x,
\end{equation}
where $\vectornorm{\cdot}_{0}$ denotes the $\ell_0$ norm. However, the $\ell_0$ minimization is known to be intractable  and a common approach is a convex relaxation based on the following $\ell_1$ minimization problem 
\begin{equation}
\underset{\x \in \real^{m}}{\min}\quad \vectornorm{x}_{1} \quad \text{s.t.} \quad \y = \A\x,
\end{equation}
where $||\cdot ||_1$ denotes the $\ell_1$ norm. CS theory shows that the $\ell_1$ relaxation exactly recovers the underlying sparse solution assuming certain conditions on $\A$ such as the restricted isometry property (RIP) or the coherence condition \citep{candes2006stable}. RIP is known to hold with high probability for random matrices and the mutual coherence can be employed for deterministic matrices \citep{donoho2003optimally,gribonval2003sparse}. 

We highlight two limitations of the standard CS. The first is the assumption of a fixed measurement matrix which is constraining in cases where an optimal predefined measurement matrix is a priori unavailable. The more general sparse coding framework learns suitable dictionaries from data adapted to the task at hand  \citep{engan2000multi,aharon2006k,elad2006image,jiang2013label}. 
Second, measurement matrices do not always satisfy RIP or coherence condition which are properties leveraged to guarantee the recovery of sparse solutions. In this paper, we illustrate these limitations in the context of the manifold learning problem. 

Sparse subspace clustering \citep{elhamifar2013sparse} is  a method to cluster data that lie on the union of manifolds. It relies on the principle of self-representation, which expresses a given data as a sparse combination of its neighbors. A recent work \citep{tankala2020manifold} utilizes dictionary learning and substitutes self representation with sparse convex combinations of dictionary columns. In this scalable approach, the size of the dictionary depends on intrinsic dimensions of data. In both frameworks, the dictionary atoms are in fact correlated and standard CS theory assuming RIP or coherence is not applicable. In addition, the $\ell_1$ regularization to promote sparsity can be geometrically oblivious. For instance, a common step in manifold learning is estimate of local geometry by reconstructing a point from nearby points \citep{roweis2000nonlinear}. With that, a sparse reconstruction of a point that uses far away dictionary atoms is sub-optimal.

\textbf{Related work}: Given the mentioned limitations of standard CS, the works in \citep{elhamifar2011sparse,tankala2020manifold} use a weighted $\ell_1$ penalty which take into account the proximity of dictionary atoms. Our work is in the spirit of weighted $\ell_1$ minimizations and structured compressive sensing \citep{khajehnejad2009weighted,vaswani2010modified,friedlander2011recovering,pilanci2012recovery}. We note that the weighted $\ell_1$ regularization employed in this paper resembles a Laplacian smoothness term \citep{dornaika2019sparse,cai2010graph}. The proposed weighted $\ell_1$ regularization in its exact form has also been used in \citep{zhong2020subspace}.

\textbf{Contributions}: We propose weighted $\ell_0$ and $\ell_1$ metrics that account for locality and allow a representation of a point using neighborhood points.
Under a certain generative model of the points, we show that the $\ell_0$ and $\ell_1$
problem are equivalent. To our knowledge, the analysis of the weighted $\ell_0$ and  weighted $\ell_1$ metrics and applications to manifold learning is a novelty
of this work.

\section{Proposed Method}

\begin{figure}[ht]
\begin{center}
    \includegraphics[scale=1]{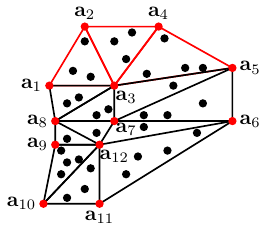}
    \caption{The red dots indicate the atoms which generate the data points. Each black dot, denoting a data point, is a convex combination of
    three atoms which are vertices of the triangle the point belongs to. We show that the optimal weighted $\ell_0$ and $\ell_1$ metric is based on a point representing itself using the vertices of the triangle it belongs to.  }
    \label{fig:illustration}
    \end{center}
\end{figure}

We consider $m$ landmark points $\a_1, \a_2, ..., \a_m$ with a unique Delaunay triangulation \citep{lee1980two}. In this setting, each point in the set $\{\y_i\}_{i=1}^{n}\in \real^{d}$ is generated from a convex combination of at most $d+ 1$ atoms. Figure \ref{fig:illustration} shows an example with $d=2$ i.e. data points in $\real^{2}$. Compactly, we have $\y_i = \A\x_i$ where the $d\times m$ matrix $\A$ is the dictionary defined as $\A = [\a_1,\a_2,...,\a_m]$  and $\x_i \in \real^{m}$
is the coefficient vector $\x_i^T = \begin{bmatrix}x_{i1}& x_{i2}& \hdots & x_{im}\end{bmatrix}$ such that $x_{ij} \ge 0$ for all $j$ and $\sum_{j=1}^{m} x_{ij} = 1$ i.e. the coefficient vector $\x_i$ 
is supported on the probability simplex in $\real^{m}$. Here on, $\Delta^{p}\equiv \{\mathbf{z} \in \real^{p}: \sum_{i=1}^{p} z_i = 1 , \mathbf{z}\ge \mathbf{0}\}$ denotes the probability simplex in $\real^{p}$. 

Before we discuss the weighted $\ell_0$ metric, we state the following definition of Delaunay triangulation that is central to our analysis.  
\begin{definition}\label{def:dela_empty_criteria}
  A Delaunay triangulation of a set of $m$ points $\mathbf{A}=\{\mathbf{a}_1,\mathbf{a}_2,...,\mathbf{a}_m\}$ in $\real^{d}$, $\text{DT}(\mathbf{A})$, is any triangulation of $\mathbf{A}$ such that for every $d$-simplex
in $\text{DT}(\mathbf{A})$, the circumscribing hypersphere of the $d$-simplex does not contain any other point of $\mathbf{A}$. 
\end{definition}
Our aforementioned goal of representing by local dictionaries  motivates the following definition of a weighted $\ell_0$ metric.
\begin{definition}
Assume $m$ landmark points $\a_1, \a_2, ..., \a_m$ in $\real^{d}$ have a unique Delaunay triangulation $\text{DT}(\A)$. Let $\z\in \real^{d}$ be an interior point of a $d$-simplex of $\text{DT}(\A)$ with circumcenter $\mathbf{c}$. The \emph{weighted $\ell_0$ norm} of any $\x\in \Delta^{m}$ is defined as
\begin{equation} \label{eq:weighted_l0_defn}
 \ell_{W,0}(\x) = \frac{1}{\|\x||_0}\sum_{i=1}^{m} \mathbf{1}_{\real_{+}}(x_i) ||\mathbf{c}-\a_i||^2,
 \end{equation}
 where $\mathbf{1}_{\real_{+}}(x_i)=1$ if $x_i>0$ and $0$ otherwise. 
\end{definition}
Given the above definition of  a weighted $\ell_0$ metric and the fact that a given point $\z$ admits different representations as a convex combination of the dictionary atoms, 
the natural question is the sense in which this metric is minimal i.e. among the different representations, which ones admit minimal values in this metric? 
The following theorem shows that the local reconstruction is minimal in the weighted $\ell_0$ metric.

\begin{theorem}
Let $\mathbf{a}_1, \ldots, \mathbf{a}_m \in \real^{d}$ be a set of points with a unique Delaunay triangulation $\text{DT}(\A)$. Let $\y \in \real^{d}$ be an interior point of the $d$-simplex of $\text{DT}(\A)$ with circumcenter $\mathbf{c}$ and radius $R$.  Assume $\y = \A\x^{*}$, where $\x^*$ has $d+1$ non-zero entries, i.e., $||\x^*||_{0}=d+1$, and the vertices corresponding to the indices of the non-zero entries of $\x^*$ are the vertices of the $d$-simplex of $\text{DT}(\A)$ that contain $\y$. Consider the following $ \ell_{W,0}$ minimization problem.
\begin{equation} \label{eq:l0_result}
 \underset{\x \in \Delta^{m}}{\min}\quad  \ell_{W,0}(\x) \quad \text{subject to} \quad \y = \A\x.
\end{equation}
The optimal solution to the above program is $\x^{*}$.
\end{theorem}

\begin{proof}
Consider a $d$-simplex of $\text{DT}(\A)$ containing $\mathbf{y}$ defined by the vertices $\{\a_j : j \in T, |T|=d+1\}$. Using vertices in $T$, $\y$ can be represented as a convex combination using coefficient vector $\x^*$. Let $\x$ be another feasible solution of the program with support $T'$. We now show that a representation using $\x^*$ is optimal:
\begin{align*}
\frac{1}{||\x||_0} \sum_{i \in T'} \mathbf{1}_{\real_{+}}(x_i) ||\mathbf{c}-\a_i||^2 
&>R^2\sum_{i \in T'} \frac{\mathbf{1}_{\real_{+}}(x_i)}{||\x||_0}\\
&=R^2\\
&= R^2 \sum_{i \in T} \frac{\mathbf{1}_{\real_{+}}(x_i^*)}{||\x^*||_0}\\
&=\frac{1}{||\x^*||_0}\sum_{i \in T}\mathbf{1}_{\real_{+}}(x_i^*)||\mathbf{c}-\a_i||^2.
\end{align*}
Above, the first inequality follows from the definition of a Delaunay triangulation (see Definition $1$). 
Therefore, the sparse representation using the vertices in $T$ 
is the optimal solution to the $\ell_{W,0}$ minimization problem. 
\end{proof}
Motivated by the fact that the weighted $\ell_0$ metric enforces local reconstructions and with the goal of obtaining a regularization amenable to optimization, we define
a convex relaxation of the weighted $\ell_0$ problem. 
\begin{definition}
Assume $m$ landmark points $\a_1, \a_2, ..., \a_m$ in $\real^{d}$ with a unique Delaunay triangulation. Let the point $\z\in \real^{d}$ have the following representation,
 $\z = \sum_{i=1}^{m} x_i \a_i$ with  $\x\in \Delta^m$. The \emph{weighted $\ell_1$} metric is defined as follows:
\begin{equation} \label{eq:weighted_l1_defn}
 \ell_{W,1}(\x) = \sum_{i=1}^{m} x_i ||\z-\a_i||^2.
 \end{equation}
\end{definition}
Analogous to CS theory, the next question is the sense in which a weighted $\ell_1$ minimization is equivalent to a weighted $\ell_0$ minimization problem. This equivalency
is summarized in the theorem below. 
\begin{theorem}
Let $\mathbf{y} \in \real^d$ lie in the convex hull of the points $\mathbf{a}_1, \ldots, \mathbf{a}_m$ in $\real^{d}$, and let $\mathbf{x}^* \in \Delta^m$ be a solution to the following optimization problem
\begin{equation} \label{eq:main_opt}
\begin{aligned}
& \underset{\mathbf{x} \in \Delta^{m}}{\text{minimize}}
& &   \ell_{W,1}(\x)\\
& \text{subject to}
& & \mathbf{y} = \sum_j x_j \mathbf{a}_j.\\
\end{aligned}
\end{equation}
If the points $\mathbf{a}_1, \ldots, \mathbf{a}_m$ have a unique Delaunay triangulation, then the set of points $\mathbf{a}_j$ such that $\mathbf{x}^*_j \neq 0$ comprise the vertices of the d-simplex of $\text{DT}(\A)$  that contains $\mathbf{y}$. 

\end{theorem}

\begin{proof}
Let $S$ denote the set of indices corresponding to the vertices of the $d$-simplex of $\text{DT}(\A)$ that contain $\y$. Since $\y$ is an interior point of this $d$-simplex, it can be written as $\y = \A\mathbf{x}^*$ where $\x^*\in \Delta^m$. Let $\x \in \Delta^m$ be another feasible solution with support $T$. We will use a proof by contradiction by assuming that $\x$ is the optimal solution. 
Observe that for any vector $\mathbf{c} \in \real^d$, the identities $\mathbf{y} = \sum_j x_j \mathbf{a}_j$ and $\sum_j x_j = 1$ imply
\begin{align*}
    & \sum_j x_j \|\mathbf{y} - \mathbf{a}_j\|^2 \\
    &= \sum_j x_j(\|\mathbf{a}_j - \mathbf{c}\|^2 - 2\langle \mathbf{y} - \mathbf{c}, \mathbf{a}_j - \mathbf{c} \rangle + \|\mathbf{y} - \mathbf{c}\|^2) \\
    &= \sum_j x_j\|\mathbf{a}_j - \mathbf{c}\|^2 - \|\mathbf{y} - \mathbf{c}\|^2.
\end{align*}
By the definition of a Delaunay triangulation, there is a circumscribing hypersphere with center $\mathbf{c}$ and radius $R$ such that $\|\mathbf{a}_j - \mathbf{c}\| = R$ for each $j \in S$ and $\|\mathbf{a}_j - \mathbf{c}\| > R$ for each $j \notin S$ (this is where we use the assumption that the triangulation is unique). Using the fact that the support of $\mathbf{x}^*$ is  $S$, we have
\begin{align*}
    \sum_{j} x'_j \|\mathbf{y} - \mathbf{a}_j\|^2 &= \sum_j x'_j \|\mathbf{a}_j - \mathbf{c}\|^2 - \|\mathbf{y} - \mathbf{c}\|^2 \\
    &< \sum_j x_j \|\mathbf{a}_j - \mathbf{c}\|^2 - \|\mathbf{y} - \mathbf{c}\|^2 \\
    &= \sum_{j} x_j \|\mathbf{y} - \mathbf{a}_j\|^2,
\end{align*}
which contradicts the optimality of $\mathbf{x}$. Therefore,  $\{\mathbf{a}_j : \mathbf{x}_j \neq 0\}$ are the vertices of the $d$-simplex of $\text{DT}(\A)$ containing $\mathbf{y}$.
\end{proof}

\section{Dictionary Learning Algorithm}

The utility of the proposed model employing weighted $\ell_1$ regularization discussed in the previous section depends on the generative dictionary. In practice,
the dictionary $\A$ is not available and needs to be estimated from the input data points. Let $\Y = [\y_1, \ldots, \y_n] \in \real^{d \times n}$ be a set of $n$ data points in $\real^d$.
For a data point $\y \in \real^{d}$, we consider the following minimization problem.
\begin{equation}\label{eq:main_opt_relax}
\underset{\A \in \real^{m\times d},\x\in \real^{m}}{\min}\quad   \frac{1}{2}\|\y - \A \x \|^2 + \lambda \sum_{j=1}^m x_{j}\|\y - \a_j\|^2.
\end{equation}  
Let  $ \L(\A, \y, \x)  =  \frac{1}{2}\|\y - \A \x \|^2 + \lambda \sum_{j=1}^m x_{j}\|\y - \a_j\|^2$ if $\x \in S$ and $ \infty$ otherwise with $S  = \{\x \in \real^m : x_1, \ldots, x_m \ge 0 \text{ and } \sum_{j=1}^m x_j = 1\}$ denoting the probability simplex in $\real^m$. We note that the parameter $\lambda$ balances the reconstruction loss with the weighted $\ell_1$ regularization
and determines the sparsity of $\x$.  

If the dictionary is fixed, the resulting problem is a weighted $\ell_1$ minimization problem and can be efficiently solved \citep{asif2012weightedl1}. Given a fixed coefficient, optimizing over the dictionary is the dictionary learning problem. We have proposed the $\mathrm{K}$-Deep Simplex (KDS) algorithm for this purpose in an earlier work \citep{tankala2020manifold}. Given data, a classical method to learn the dictionary alternates between sparse approximation and dictionary update step \citep{agarwal2016altmin}. KDS utilizes this along with algorithm unrolling \citep{eldar2019unrolling, gregor2010learning, hershey2014unfolding}, a framework to recast a structured optimization framework into a neural network, to solve the optimization  problem in \eqref{eq:main_opt_relax}.  We develop an autoencoder architecture to solve the optimization problem. We summarize the main steps of the KDS algorithm below and refer the interested reader to \citep{tankala2020manifold} for details. 

\textbf{Encoder:} We employ the accelerated projected gradient descent \citep{bubeck2015convex} to compute the optimal coefficient for  
\begin{equation}
\x^*(\A, \y) \in \argmin_{\x} \L(\A, \y, \x).
\end{equation}
Starting with initialization $\x^{(0)} = \tilde{\x}^{(0)} = \mathbf{0}$, the algorithm updates can be written as follows
\begin{align}
\x^{(t+1)} &= \mathcal{P}_S\left(\tilde{\x}^{(t)} - \alpha \nabla_{\x} \L(\A, \y, \tilde{\x}^{(t)})\right) \\
\tilde{\x}^{(t+1)} &= \x^{(t + 1)} + \gamma^{(t)}(\x^{(t + 1)} - \x^{(t)}),
\end{align}
for $0 \le t \le T$. The parameter $\alpha$ is a step size, and the constants $\gamma^{(t)}$ are given by the recurrence \begin{equation}\eta^{(0)} = 0,\quad \eta^{(t+1)} = \frac{1+\sqrt{1+4\eta^{(t)}}}{2},\quad \gamma^{(t)} = \frac{\eta^{(t)} - 1}{\eta^{(t+1)}}.\end{equation} The gradient of the loss function $\L$ is given by
\begin{equation}
\nabla_{\x} \L(\A, \y, \x) = \A^\top(\A\x - \y) + \lambda \sum_{j=1}^m \|\y - \a_j\|^2 \e_j.
\end{equation}
The operator $\mathcal{P}_S$ projects onto $S$, the probability simplex and has the following form \citep{wang2013projection}  \begin{equation}\mathcal{P}_S(\x) = \mathrm{ReLU}(\x + b(\x)\cdot \mathbf{1}),\end{equation} with $b : \real^m \to \real$ denoting a piecewise-linear bias function. In summary, the approximate optimal code $\x^{(T)}(\A, \y) \approx \x^*(\A, \y)$ is an output of a recurrent encoder with input $\y$, weights $\A$, and activation function $\mathcal{P}_S$.

\textbf{Decoder and Backward Pass :} Given $\A$ and $\x$, the decoder approximately reconstructs the input $\y$ by computing $\hat\y = \A\x^{(T)}$.
The final step it to find the optimal network weights by solving $\displaystyle \label{eq:empirical_loss_for_ae} \A^* \in \argmin_{\A} \frac{1}{n}\sum_{i=1}^n \L(\A, \y_i, \x^*(\A, \y_i))$. We minimize $\displaystyle \frac{1}{n}\sum_{i=1}^n \L(\A, \y_i, \x^{(T)}(\A, \y_i))$ by backpropagation through the autoencoder.  An advantage of this optimization framework is that the encoding and decoding steps are amenable to GPU parallelizations across the data points. 

\section{Numerical Experiments}

\begin{figure}
\centering
    \includegraphics[width=0.24\linewidth]{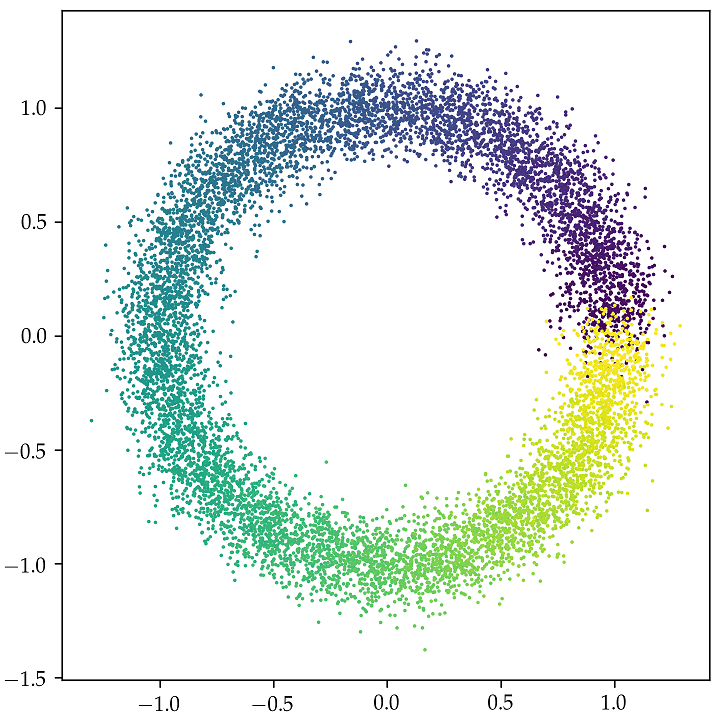}
    \includegraphics[width=0.24\linewidth]{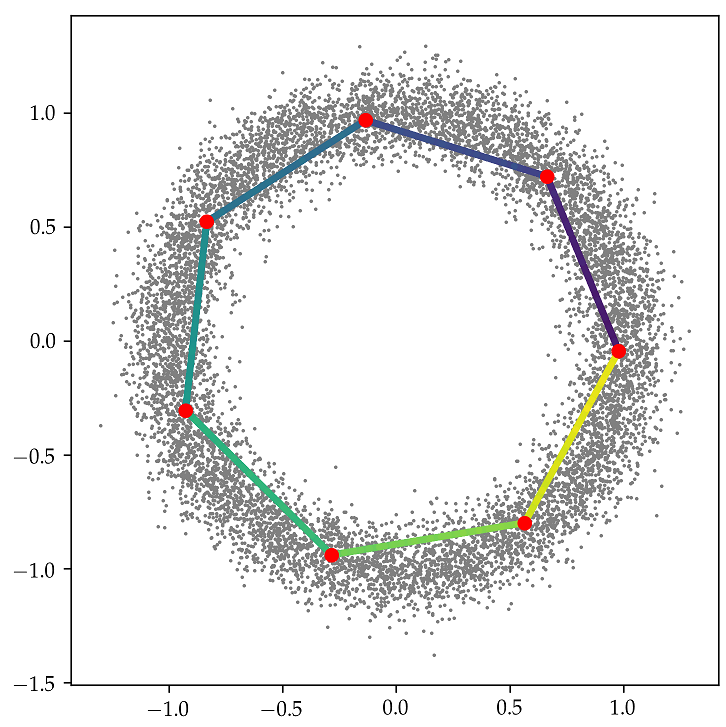} 
    \includegraphics[width=0.24\linewidth]{input_cropped.png}
    \includegraphics[width=0.24\linewidth]{output_cropped.png}
    \caption{Circle and two moons. Autoencoder input (first and third) and output (second and fourth), with learned atoms marked in red.}
    \label{fig:synthetic}
\end{figure}

In this section, we show that the proposed regularization is useful for the manifold learning task. We first consider one-dimensional manifolds in $\real^{2}$. Figure \ref{fig:synthetic} shows two such data sets.  The first data is a unit circle in $\real^{2}$. The second data we consider is the classic two moon data set  \citep{ng2001spectral}. The latter dataset consists of two disjoint semicircular arcs. To test the robustness of the proposed regularization and the KDS algorithm, a small Gaussian white noise is added to each data point.  Figure \ref{fig:synthetic} shows the results of training the autoencoder on these data sets. Using the weighted $\ell_1$ metric and employing the KDS algorithm, each data point is accurately represented as sparse convex combinations of neighborhood atoms. In addition, the
learned atoms are geometrically significant and can be interpreted as the dictionary that generate the points up to additive noise. 

\subsection{Application: Clustering} The learned sparse coefficients can be used for downstream tasks. Here, we illustrate their use for the clustering task and compare our results to competitive clustering methods.
The first is the $k$-means (KM) algorithm \citep{lloyd1982least} which is a centroid based clustering algorithm. The second is the sparse manifold clustering and embedding (SMCE) algorithm proposed in \citep{elhamifar2011sparse}. Therein, the authors also consider a proximity regularization. However, the dictionary is constituted from all data points. In contrast, our method learns few atoms from
the input data. We evaluate the algorithms on two datasets. The first is a noisy two moon dataset. The second is the MNIST dataset \citep{lecun1998gradient} which is a database of $10$ different digits each represented as  a $28 \times 28$ grayscale image. We consider a subset of the data by considering $5$ digits, $0,3,4,6$ and $7$. For KDS and SMCE, we run spectral clustering \citep{ng2001spectral}
on a similarity graph derived from the coefficients. Table \ref{tab:acc} summarizes the results.  We note that the proposed algorithm achieves the best accuracy over the baselines. As important as the accuracy is the fact that the clustering is performed on a sparse similarity graph. This ensures that KDS is fast and makes it an efficient scalable algorithm for large datasets. 
\begin{table}[h!]
    \centering
    \begin{tabular}{|c|c|c|c|c|}
    \hline
        Method & Moons & MNIST-$5$  \\
        \hline
        KM       & 0.75 & 0.91       \\
        SMCE   & 0.88   & 0.94   \\
        KDS     &    $\mathbf{1.0}$ & $\mathbf{0.99}$       \\
        \hline
    \end{tabular}
    \caption{Clustering accuracy for various data sets. Accuracy is defined the percentage of correct matches with respect to the ground truth labels of the data.  }
    \label{tab:acc}
\end{table}
 
 \section{Conclusion}
 
 In this work, we propose weighted $\ell_0$ and weighted $\ell_1$ regularizations that promote the representation  of a data point using nearby dictionary atoms. Assuming that data points are generated from a convex  combination of atoms, represented as vertices of a unique Delaunay triangulation,  we prove that the weighted $\ell_1$ regularization recovers the underlying sparse solution. In the general setting, we discuss an efficient algorithm  to learn the sparse coefficients and dictionary atoms. In the context of the manifold learning problem, our experiments show that the proposed regularization obtains a geometrically meaningful estimate of local  geometry. We test the algorithm on the clustering problem and show that the proposed framework is efficient and yields accurate results.

\bibliographystyle{IEEEtranN}
\interlinepenalty=10000
\bibliography{Weighted_l1}

\end{document}